\documentclass[12pt]{article}
\usepackage{pdproc,epsfig} 

\begin{document}

\renewcommand{\thefootnote}{\alph{footnote}}
  
\title{
RESULTS FROM MINIBOONE
}

\author{ BYRON P. ROE\\ For the MiniBooNE Collaboration}

\address{ Department of Physics, University of Michigan
450 Church Street\\
Ann Arbor, MI 48109-1040, U.S.A. \\
 {\rm E-mail: byronroe@umich.edu}}




\abstract{Recent results from MiniBooNE are described.  These include
neutrino oscillation results, low energy anomaly,  
and neutrino/antineutrino cross sections.
}
   
\normalsize\baselineskip=15pt

\section{The MiniBooNE Experiment}
MiniBooNE was proposed in Summer 1997 to examine the result obtained in 
the LSND experiment.  MiniBooNE has been running since 2002.  The LSND
experiment\cite{lsnd}
presented a 3.8 $\sigma$ signal 
for $\bar{\nu}_\mu\rightarrow \bar{\nu}_e$ oscillations
with $\Delta m^2$ of the order of 1 eV$^2$ and $\sin^22\theta=0.26\pm0.08$.
This result is incompatible with the results obtained for solar and
atmospheric neutrinos in the framework of the three-neutrino standard model.

MiniBooNE makes neutrinos using the Fermilab Booster 8 GeV proton beam
incident on a 71 cm long beryllium target inside of a toroidal focusing horn.
Up to $4\times10^{12}$ protons are contained within a $\sim 1.6\ \mu$s
beam spill at a rate of up to 4 Hz.  Positively charged pions and kaons
are focused into a 50 m long, 91 cm radius decay region. (See Figure 1.)
$L/E$ for MiniBooNE is quite similar to that for LSND, but the MiniBooNE
mean neutrino beam energy of about 800 MeV is far greater than the LSND
energy of about 50 MeV, resulting in very different systematics for the
two experiments.

The first MiniBooNE oscillation result was based on an exposure of  
5.58$\times 10^{20}$ protons on target.  Most of the neutrinos produced
are $\nu_\mu$, but there is an intrinsic $\nu_e$ fraction of about 0.5\%,
which comes from kaon and muon decay.  About 39\% of the neutrino interactions
in the detector are charged-current quasi-elastic (CCQE) scattering, 16\% are
neutral-current (NC) elastic scattering, 29\% are charged-current (CC) single
pion production and 12\% are NC single pion production.

The neutrino detector is located 541 m downstream of the beryllium target and
is 1.9 m above the center of the beam line.  The detector consists of a
spherical tank with 610 cm inner radius containing 800 tons of pure mineral
oil (CH$_2$) of density 0.86 g/cm$^3$ and index of refraction of 1.47.
Fast charged particles produce both prompt, directional Cherenkov light and
longer time constant scintillation light with a ratio of about 8:1.
However, because of significant absorption and fluorescence, the ratio
of prompt to delayed light at the detector photo-tubes is about 3:1. 
  The detector consists of an inner spherical target region
of radius 575 cm with 1280 equally-spaced inner-facing 8-inch photomultiplier
tubes 
(PMT) providing 10\% photocathode coverage.  There is an optically 
isolated outer veto shield region 35 cm thick with 240 8-inch PMTs.
The detector has been designed to detect and measure neutrino events in
the energy range from 100 MeV to a few GeV.  The event vertex, 
outgoing particle directions and
visible energies are determined by event reconstruction, which also 
determines parameters allowing separation of $\nu_e$ and $\nu_\mu$
events.

\begin{figure}
\vspace*{13pt}
         \centering{\mbox{\epsfig{figure=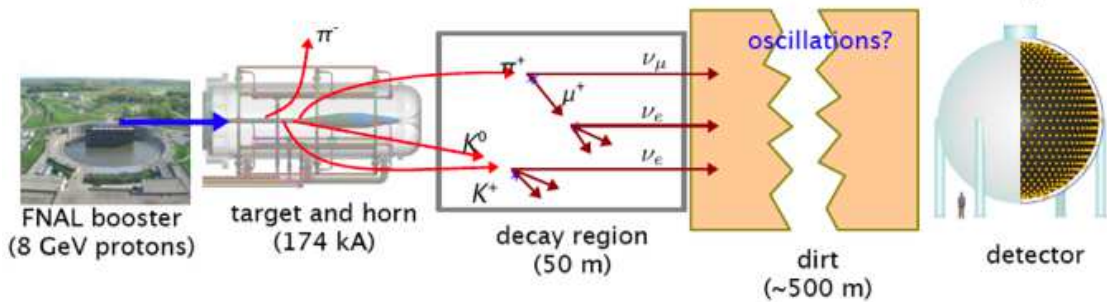,width=15.0cm}}}
\vspace*{0.2truein}		
\caption{MiniBooNE beam}
\label{fig:mbbeam}
\end{figure}
\section{Oscillation Analysis}
In the initial MiniBooNE oscillation publication\cite{mbosc},
two separate, but quite compatible, results
coming from two different reconstruction-particle identification
packages were presented.  No LSND-like signal was seen.  
A first attempt to combine 
these two results to produce better limits has now been made
and is shown in Figure 2. Below $\Delta m^2$ of 1 eV$^2$, the  90\% CL limit
is improved by 10-30\%.  This result used CCQE events above 475 MeV.
\begin{figure}
\vspace*{13pt}
         \centering{\mbox{\epsfig{figure=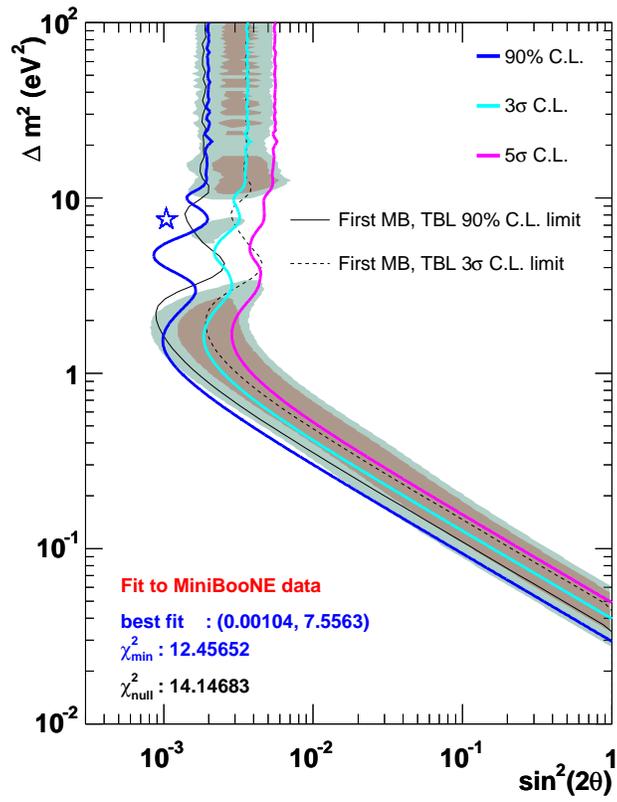,width=8.0cm}}}
\vspace*{0.2truein}		
\caption{New MiniBooNE $\nu_\mu$-$\nu_e$ oscillation limits 
combining the two independent 
reconstruction-particle identification methods.}
\label{fig:osclim}
\end{figure}

Global simultaneous fits to LSND, MiniBooNE, KARMEN2\cite{karmen} and
Bugey\cite{bugey} have been performed\cite{global}.  
The results of each experiment were
converted to a $\chi^2$.  However, only $\Delta \chi^2$ was available;
the goodness of fit of individual experiments was not available for each
case.  Two-dimensional fits varying both oscillation parameters and
one-dimensional fits, varying only $\sin^22\theta$ were done.  The fits
were done using the prescription of Maltoni and Schwetz\cite{maltoni}.
For each $\Delta m^2$, the
one-dimensional fit answers the question, ``If this is the true
$\Delta m^2$, what is the compatibility?''  The two-dimensional
fits are presented in Table 1.  Using all four experiments, the 
maximum compatibility is 3.9\%.  If KARMEN2 is omitted, the
maximum compatibility drops to 2.1\%; the KARMEN2 experiment has
the effect of diluting the results.  If LSND is omitted, the
maximum compatibility of the other three experiments is good, 25.4\%.
The limits of the Bugey experiment in the region below $\Delta m^2=1$
eV$^2$ are quite important for the global fit.  The limits from the 
one-dimensional fits are quite comparable to those from the two-dimensional
fits. 
\begin{table}[h]
\caption{Maximum Compatibility for 2-D Global Fits to Experiments. 
The X
indicates which experiments were included in the analysis.
}\label{tab:exp}
  \small
  \begin{tabular}{||c|c|c|c|c|c|l||}\hline\hline
  {} &{} &{} &{} &{} &{} &{}\\
  LSND & KARMEN2 & MB & Bugey & Max. Comp. (\%) & $\Delta m^2$ & $\sin^22\theta$ \\
  {} &{} &{} &{} &{} &{} &{}\\
  \hline
  {} &{} &{} &{} &{} &{} &{}\\
X & X & X & & 25.4 & 0.072 & 0.26\\
X & X & X & X & 3.94 & 0.24 & 0.023\\
  \hline
  {} &{} &{} &{} &{} &{} &{}\\
X & & X & & 16.0 & 0.072 & 0.26\\
X & & X & X & 2.1 & 0.25 & 0.023\\
\hline
  {} &{} &{} &{} &{} &{} &{}\\
 & X & X & & 73.4 & 0.052 & 0.15\\
 & X & X & X & 27.4 & 0.22 & 0.012\\
  \hline\hline
\end{tabular} 
\end{table}

\section{Low Energy Anomaly}

In the MiniBooNE oscillation paper\cite{mbosc}, 
the $\nu_e$ CCQE energy spectrum from 300 to 475 MeV was found to have an 
excess of $96\pm 17\pm 20$ events above
that predicted from the null oscillation fit.  No oscillation parameters
fit the entire spectrum.  Parameters which come close to fitting the low energy
data predict too many high energy events. 
 A
number of studies of these low energy events are ongoing; it is hoped that a
complete update will be available in Summer 2008.  A status
report on these studies is given below.  None of the effects examined
are expected to have any appreciable effect above 475 MeV.

The analysis has been
extended down to the $\nu_e$ CCQE energy region of 200-300 MeV, 
where an anomaly is also seen, similar in size to that in the 300-475 MeV
region. 
Above 475 MeV there is essentially no excess, $22\pm19\pm35$ events. 
\begin{itemize}
\item{} No evidence of instrumental background has been found.  Spatial
and angular 
distributions of events within the detector are as expected as are
event timings.  Energy
calibrations are done with data as described in the initial 
oscillation publication.
\item{} The two reconstruction, particle identification packages give
consistent analyses.
\item{} Particle identification studies indicate that the excess is
completely consistent with being electrons or gamma rays. (MiniBooNE
cannot distinguish an electron track from the electron-positron
pair resulting from gamma ray conversion.)
\item{} NC events producing an excited nucleon $\Delta$, usually decay
by $\pi^0$-nucleon, but occasionally by $\gamma$-nucleon, a 
$\Delta$-radiative decay.  This latter
decay emulates a $\nu_e$ CCQE event. The  $\pi^0$-nucleon events have
been measured in the detector.  This measurement calibrates the number of 
$\Delta$-radiative decays since the branching fraction for the mode
is known and corrections for nuclear and threshold effects can be
calculated.
\item{} Neutrino events occurring in material outside the detector can produce
$\pi^0$'s.  If one of the decay gamma rays enters the detector it
can simulate a $\nu_e$ CCQE event.  The gamma rays tend to be of low energy.
Since the gamma conversion length
is short, 70 cm, the extrapolated path length back to the detector
wall tends to be short.  By
selecting events with short extrapolated length, these events can
be enhanced and the size of the effect calibrated. 
\item{} An energy dependent cut on the backward extrapolated path length
of the events is now being implemented.  This has the effect of reducing
the number of ``dirt'' events with only minor loss of signal.
\item{} Various effects can cause events with an outgoing muon or
pion to look like  $\nu_e$ CCQE events.  For example, a
preprint\cite{bodek} suggested that muon bremsstrahlung might be a
cause of the anomaly.  However, for 83\% of the events with an
outgoing muon, the muon decay is detected.  These
events appear as two sub-events with a time delay that of the muon
decay time.  By eliminating the muon decay sub-event, a sample of
guaranteed muons or pions is obtained.  These events are then
reprocessed through the reconstruction and particle identification
steps to see whether any pass the  $\nu_e$ CCQE criteria.  The fraction
of the anomaly due to these events is about 2\%\cite{mubrem}.
\item{} If one of the gamma rays from $\pi^0$ decay is absorbed by
means of a photo-nuclear process, the remaining gamma ray may appear
to be a $\nu_e$ CCQE event.  Photo-nuclear processes were not included
in the version of GEANT originally used.  Using data from other
experiments, the cross section for this process as a function
of energy was estimated as was the effect of extra final state particles.
This process reduces the anomaly.  The size of the reduction and the
systematic errors on it are still being calculated.
\item{} A more comprehensive study of hadronic errors and a better
handling of $\pi^\pm$ interactions is ongoing.  It is expected that
this will result in a reduction of the anomaly.
\item{} A better handling of the $\pi^0$ background calculation is
underway.  This will reduce the anomaly.
\item{} Improved measurements of neutrino induced $\pi^0$'s is expected
to increase the anomaly.
\item{} An improvement in the handling of beam $\pi^+$ production
uncertainties is in the process of being implemented.  The effect on
the anomaly is uncertain.
\item{} Examination of the anomaly in $\bar{\nu}$ events will be done
shortly.  The fractions of various kinds of background in both samples 
is similar, except that the $\bar{\nu}$ beam events have a larger fraction
of intrinsic $\nu_e$ events (2.45\% vs 0.5\% for events in the
$\nu$ beam).  Different hypotheses for the excess, however,
can have measurably different effects in the two samples.
\end{itemize}

\subsection{Some Theoretical Suggestions for Causes of the Anomaly}
One suggestion is a new process within the standard model, an axial
anomaly\cite{hhh}.
This suggestion involves a triangle diagram.  A nucleon emits an $\omega$
which goes to one vertex of the triangle.  The neutrino emits a $Z$
which goes to the second vertex of the triangle, and a gamma ray is
emitted from the third vertex.  The low energy limit cross section with
no nucleon recoil is
$$\sigma = {\alpha g_\omega^4G_F^2\over 480\pi^6m_\omega^4}E^6_\nu=
2.2\times 10^{-41}\left({E_\nu\over{\rm GeV}}\right)^6\left({g_\omega\over 10}
\right)^4.\eqno{1}$$
The cross section is expected to saturate at higher energies.  A more 
detailed calculation of the expected gamma ray energy and angular 
distribution using a Monte Carlo calculation from the complete amplitude is
now underway.  Normalization is a problem since $g_\omega$ can vary by a
factor of about three, from 10 to 30, and it appears in the cross section
to the fourth power.

Another suggestion\cite{paraphoton}
involves a new light gauge boson called a paraphoton.  There is an
MSW-like potential in matter which affects low energy neutrino 
oscillations.  It makes LSND and MiniBooNE results compatible and obtains
a low energy anomaly about 40\% of that seen in MiniBooNE. The paraphoton
has a mass of $\sim10$ KeV.  (It needs to be short range to avoid fifth
force measurements).  It has a very low coupling strength to B$-$L with
$g^2/e^2\sim 10^{-9}$, which was thought to make it undetectable.
However, MiniBooNE has about $6\times 10^{20}$ protons on target.  Two
back-of-the-envelope calculations indicate that there might be 10-20
events in the very forward direction, looking like an anomalous $\nu_e$-$e$
scattering cross section.  A paraphoton would be produced by hadronic
bremsstrahlung of the incident proton beam and seen in the detector by
producing an electron-positron pair.  Examination of our present data is
underway.  More events will be needed for a definitive answer, but for
this process, the $\bar{\nu}$ beam is as good as the $\nu$ beam.  
This calculation indicates that MiniBooNE
is capable of doing very sensitive searches for a variety of rare processes.

\section{Neutrino and Anti-Neutrino Cross Sections}
\subsection{$\nu_\mu$ CCQE Cross Section}
The $Q^2$ dependence of the $\nu_\mu$ CCQE events has been 
fit\cite{mbccqe}
 using
a relativistic Fermi gas model. 193,709 events pass the MiniBooNE 
$\nu_\mu$ CCQE
criteria.
The binding energy and the Fermi momentum were taken from electron scattering 
data and the effective axial mass $M_A^{\rm eff}$ and Pauli blocking
parameter $\kappa$ were fitted. $M_A^{\rm eff}=1.23\pm 0.20$ GeV and
$\kappa = 1.019\pm 0.011$.  $M_A^{\rm eff}$ is larger than the values
found in previous, lower statistics, experiments, but is in agreement
with the K2K result\cite{kk}.  These are all effective values
since the Fermi gas model is a poor approximation; a better calculation is
needed.  The $Q^2$ distribution is shown in Figure 3.
\begin{figure}
\vspace*{13pt}
         \centering{\mbox{\epsfig{figure=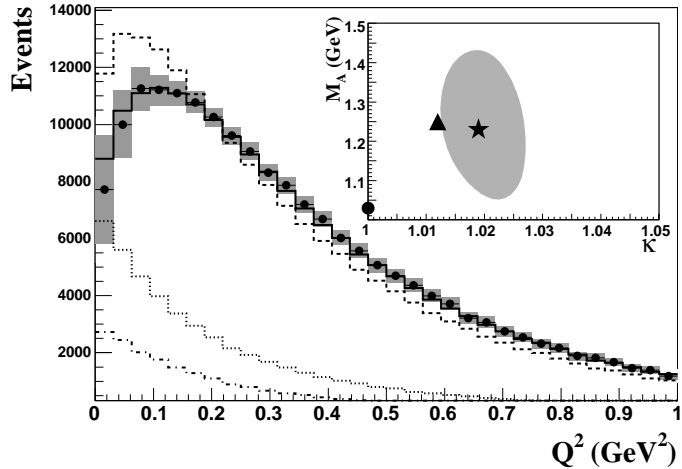,width=10.0cm}}}
\vspace*{0.2truein}		
\caption{Reconstructed $Q^2$ for $\nu_\mu$ CCQE events including systematic
errors.  The simulation before (dashed) and after (solid) fitting, is 
normalized to the data.  The dotted (dot-dash) curve shows backgrounds that
are not CCQE (``CCQE-like'').  The inset shows the $1\sigma$ contour for the
best-fit parameters (star), along with the starting values (circle), and
the fit results after varying the background (triangle).}
\label{fig:piccqe}
\end{figure}

\subsection{$\nu_\mu$ NC Elastic Cross Section}
The present results are from a 10\% sample of our data.  The flux
integrated cross section is $8.8\pm 0.6{\rm (stat)}\pm{\rm 2.0(syst)}\times
10^{-40}$ cm$^2$, and the measured axial mass is $1.34^{+0.38}_{-0.25}$ GeV.
The full data set is now under analysis.
\subsection{NC and CC $\pi^0$ Events}
It is important to calibrate the NC $\pi^0$ event sample.  
If one $\pi^0$ decay gamma rays is lost, the remaining gamma ray is
a background for our oscillation analysis.  Radiative nucleon
resonance decays form another background to $\nu_e$ events.  The NC
$\pi^0$ data events were compared with Monte Carlo predictions.  The
Monte Carlo cross sections were corrected by re-weighting Monte Carlo events
to match the data.  This was important both for the MiniBooNE oscillation
analysis and for understanding the low energy anomaly.  Figure 4 shows
a comparison of unsmeared data events and the unmodified Monte Carlo
events as a function of $\pi^0$ momentum as well as the needed
re-weighting factor.
\begin{figure}
\vspace*{13pt}
         \centering{\mbox{\epsfig{figure=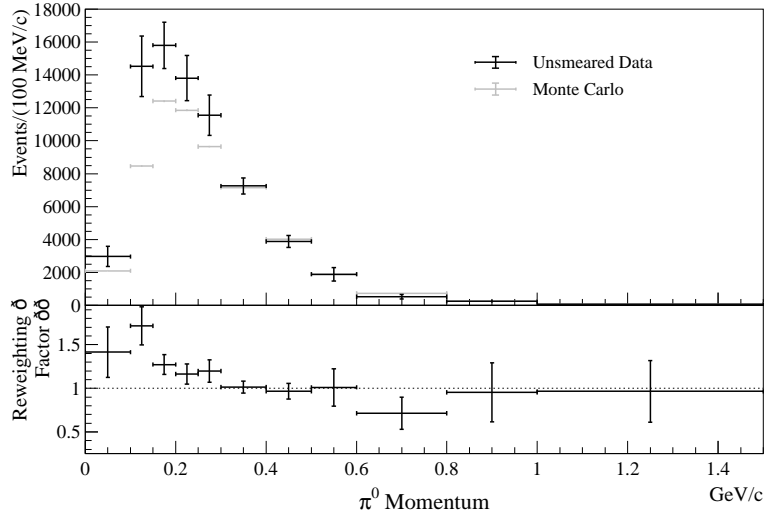,width=10.0cm}}}
\vspace*{0.2truein}		
\caption{NC $\pi^0$ Reconstruction Unsmearing. Top: Results of the $\pi^0$
unsmearing in bins of momentum.  Dark points are unsmeared data, 
light points are 
uncorrected Monte Carlo.  Bottom: Re-weighting function.}
\label{fig:pi0unsmear}
\end{figure}

This NC event sample can also be examined to find the fraction of events
not going through a nucleon resonance, but going coherently.  This
fraction was measured to be $19.5\pm 1.1{\rm(stat.)}\pm 2.5{\rm(syst.)}$\%, 
which is considerably
below the 30\% prediction of Rein and Sehgal\cite{rs} as implemented
in the NUANCE Monte Carlo\cite{nuance}.  

The CC event sample
has a gamma-gamma effective mass distribution in good agreement with
Monte Carlo expectations.  This sample of events has no coherent
contribution.

\subsection{$\bar{\nu}$ NC $\pi^0$ Cross Section}
The data and Monte Carlo agree very well in the shape of the gamma-gamma
effective mass.  When a fit is made for resonant, coherent, and background
fractions, the coherent fraction is again considerably less than predicted
by Rein and Sehgal.

\subsection{$\nu_\mu$ CC $\pi^+$ Events from the $\bar{\nu}$ Beam}
This cross section provides a direct measurement of the rate and
energy dependence of $\nu$ backgrounds in the $\bar{\nu}$ beam.
In addition, most of the $\nu$ production comes from decays of very
forward $\pi^+$, since, if they enter the focusing horn, they are
bent away from the detector.  This allows a check of the $\pi^+$ 
production at angles below that at which data exists.  Results are
shown in Figure 5.
\begin{figure}
\vspace*{13pt}
         \centering{\mbox{\epsfig{figure=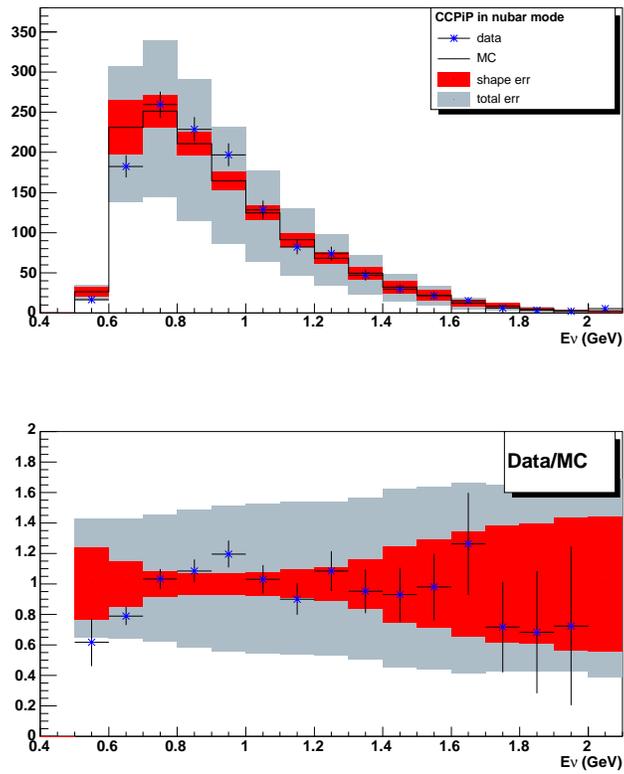,width=10.0cm}}}
\vspace*{0.2truein}		
\caption{Top:  Data (dots) and Monte Carlo (solid line) for  $\nu_\mu$ 
events giving a $\pi^+$ seen in the $\bar{\nu}$ beam. The bars indicate shape 
and total errors.  Bottom:  Ratio of data to Monte Carlo.}
\label{fig:nuinnubar}
\end{figure}


\end{document}